\documentclass[twocolumn,prl,8pt]{revtex4}
\pdfoutput=0
\usepackage{graphicx}
\usepackage{graphics}
\usepackage{floatflt}
\usepackage{eso-pic}
\usepackage{type1cm}

\graphicspath{{figures/}{figures2/}}
\DeclareGraphicsExtensions{.eps,.png}

 \newcommand{\ket}[1]{$| #1 \rangle$}

\newcommand{\phicjj}{$\Phi^x_{\rm{ccjj}}$ }
\newcommand{\deltaq}{$\Delta$} 
\newcommand{\ip}{$I_p$ }

\begin{document}
\title{Observation of Co-tunneling in Pairs of Coupled Flux Qubits}
\author{T.~Lanting}
\email{tlanting@dwavesys.com}
\author{R.~Harris}
\author{J.~Johansson}
\author{M.H.S.~Amin}
\author{A.J.~Berkley}
\author{S.~Gildert}
\author{M.W.~Johnson}
\author{P.~Bunyk}
\author{E.~Tolkacheva}
\author{E.~Ladizinsky}
\author{N.~Ladizinsky}
\author{T.~Oh}
\author{I.~Perminov}
\author{E.M.~Chapple}
\author{C.~Enderud}
\author{C.~Rich}
\author{B.~Wilson}
\author{M.C.~Thom}
\author{S.~Uchaikin}
\author{G.~Rose}


\affiliation{D-Wave Systems Inc., 100-4401 Still Creek Drive, Burnaby, B.C., V5C 6G9, Canada}

\date{\today}

\begin{abstract}
We report measurements of macroscopic resonant tunneling between the two lowest energy states of a pair of magnetically coupled rf-SQUID flux qubits. This technique provides a direct means of observing two-qubit dynamics and a probe of the environment coupled to the pair of qubits. Measurements of the tunneling rate as a function of qubit flux bias show a Gaussian line shape that is well matched to theoretical predictions. Moreover, the peak widths indicate that each qubit is coupled to a local environment whose fluctuations are uncorrelated with that of the other qubit.
\end{abstract}

\maketitle

Superconducting circuits have played an essential role in realizing quantum mechanical phenomena in macroscopic systems. One such example is the observation of macroscopic resonant tunneling (MRT) of magnetic flux between the lowest energy states of single rf-SQUID flux qubits, 
as demonstrated by several groups \cite{rouse-mrt-1995,harris-mrt-2008-truncated,bennett-decoherence-2008-truncated,crankshaw-mrt-2004-truncated}. These measurements provide both a clear signature of quantum mechanics in a macroscopic circuit at a finite temperature and in the presence of noise and a direct means of determining the tunneling energy between states. Theoretical descriptions of the MRT rate have been presented \cite{averin-mrt-2000,amin-mrt-2008} and indicate a direct connection between the profile of the MRT rate peaks and properties of the environment. Analogous measurements of the tunneling of magnetization in nanomagnets~\cite{friedman-quantummagnetism-1996,thomas-quantummagnetism-1996-truncated} suggest that MRT is responsible for dynamics in these materials as well.

In this work, we extend measurements of MRT to inductively coupled pairs of flux qubits.  We present experimental observations of tunneling between the two lowest energy states of the coupled system for several coupling strengths. These data yield two-qubit energy gaps that match those predicted by the independently calibrated Hamiltonian of the coupled system. Moreover, measurements of the two-qubit energy gap are used to infer single qubit energy gaps at $\sim h \times 10^9$ Hz without the use of microwave lines. Finally, the profile of the MRT rate versus qubit flux bias has a Gaussian lineshape with a width that is a factor of $\sqrt{2}$ larger than that of a single qubit. We argue that this observation indicates that the environments coupled to each qubit are uncorrelated.

For a single flux qubit, an MRT experiment consists of measuring the rate of tunneling of flux between two wells of the double-well potential of the rf SQUID when the lowest energy levels of each well are closely aligned. Restricting the dynamics of the single rf SQUID to its two lowest energy states allows one to map the physics of this device onto the canonical qubit Hamiltonian:
\begin{equation}
H_q = -\frac{1}{2}[\epsilon \sigma_z + \Delta \sigma_x] +
\frac{1}{2}Q\sigma_z\;,
\label{EQN:singlequbittruncatedhamiltonian}
\end{equation}
where $\sigma_{x,z}$ are Pauli matrices, $\epsilon \equiv 2I_p(\Phi^x_q-\Phi^x_0)$ is the energy
difference between the two wells, $I_p$ is the amplitude of the persistent current in the rf-SQUID loop, $\Delta$ is the tunneling energy, and $Q$ is an operator that acts on an environment that generates flux noise in the qubit. Here, $\Phi^x_q$ represents the external flux bias applied to the rf-SQUID loop and $\Phi^0_q$ is the degeneracy point. Hamiltonian~(\ref{EQN:singlequbittruncatedhamiltonian}) is valid when $|\epsilon|, \Delta \ll \hbar\omega_p$, where $\hbar\omega_p$ is the energy spacing to the next excited state of the rf SQUID. For a non-Markovian environment~\cite{weiss-dissipative-book}, the initial tunneling rate from \ket{0} to \ket{1} (eigenstates of $\sigma_z$) versus $\epsilon$ has a Gaussian profile, as given by Eq. (2) in Ref.~\cite{harris-mrt-2008-truncated}.


A natural extension to the single qubit MRT experiment is to add a second qubit that is inductively
coupled to a first qubit via a mutual inductance $M_{\rm{eff}}$. The system then has the following low energy Hamiltonian:
\begin{equation}
H_{2q} = -\frac{1}{2} \sum_{i=1}^2 \left[(\epsilon_i + Q_i)\sigma_z^{(i)} + \Delta_i\sigma_x^{(i)}\right] + J\sigma_z^{(1)}\sigma_z^{(2)}\;,
\end{equation}
where $J \equiv M_{\rm{eff}} I_{p1}I_{p2}$ is the coupling energy and all qubit-specific quantities are labeled with $i \in [1,2]$. The qubits are ferromagnetically (FM) coupled when $J< 0$. For $|J| \gg
|\epsilon_i|,\Delta_i$, the two lowest energy eigenstates are
approximately superpositions of the FM ordered states
\ket{00} and \ket{11} (eigenstates of $\sigma_z^{(1)}\sigma_z^{(2)}$). One can therefore write a two-state
Hamiltonian to describe the low energy dynamics in this subspace:
 \begin{equation}
 H_{2q} \approx -\frac{1}{2}[(\epsilon_1 + \epsilon_2) \tau_z + g \tau_x] \\
 + \frac{1}{2}(Q_1+Q_2)\tau_z\;,
\label{EQN:2qhamiltonian}
 \end{equation}
where $\tau_{x,z}$ are Pauli matrices in the above 2-dimensional
subspace and $g$ is the two-qubit energy gap:
\begin{equation}
g = \sqrt{J^2 + \frac{1}{4}(\Delta_1 + \Delta_2)^2} - \sqrt{J^2 + \frac{1}{4}(\Delta_1 - \Delta_2)^2}\;.
\label{EQN:fullanalyticgap}
\end{equation}
For the regime $\Delta_1$, $\Delta_2 \ll 2|J|$, Eq.~(\ref{EQN:fullanalyticgap}) simplifies to $g \approx \Delta_1\Delta_2/2J$.
If  $2|J| \ll \hbar\omega_p$, the nearest excited states outside of this subspace are formed from the antiferromagnetically ordered states \ket{01} and \ket{10}. If $2|J| \gtrsim \hbar\omega_p$, additional levels from the two rf SQUIDs need to be included and $g$ must be evaluated numerically. For all measurements reported herein, $2|J| \ll \hbar\omega_p$, thus justifying our use of Eqs. (\ref{EQN:2qhamiltonian}) and (\ref{EQN:fullanalyticgap}).

By adapting the derivation in Ref.~\cite{amin-mrt-2008} to the subspace spanned by \ket{00} and \ket{11}, we derive a functional form for the two-qubit co-tunneling rate from \ket{00} to \ket{11}:
\begin{eqnarray}
\Gamma_{00 \rightarrow 11}(\epsilon') =
\frac{1}{\hbar}\sqrt{\frac{\pi}{8}}\frac{g^2}{W_{2q}}\exp\left[-\frac{(\epsilon'-\epsilon_{p,2q})^2}{2W_{2q}^2}\right]\;,
\label{EQN:2q-mrt-eqn}
\end{eqnarray}
where $\epsilon' \equiv \epsilon_1 + \epsilon_2$, $\epsilon_{p,2q}$ and $W_{2q}$ represent the displacement and width of a Gaussian peak, respectively, and $\Gamma_{11 \rightarrow 00}(\epsilon') = \Gamma_{00 \rightarrow
11}(-\epsilon')$. We define the noise spectral density $S_{2q}(\omega)$ for the coupled system as follows:
\begin{equation}
S_{2q}(\omega) \equiv \int dt \, e^{i\omega t}\left<\left[Q_1(t)+Q_2(t)\right]\left[Q_1(0)+Q_2(0)\right]\right>\;,
\label{EQN:2qspectraldensity}
\end{equation}
where $\langle\ldots\rangle$ denotes averaging over all environmental modes. We use Eq.~(\ref{EQN:2qspectraldensity}) to calculate $W_{2q}$ as in Refs.~\cite{amin-mrt-2008,harris-mrt-2008-truncated}: 
\begin{eqnarray}
W_{2q}^2&\equiv&\int\! \frac{d\omega}{2\pi} \, S_{2q}(\omega) \nonumber\\
  & = & 2W^2 + \int\! \frac{d\omega}{2\pi} \, \int\! dt \, e^{i\omega t}\left<Q_1(t)Q_2(0)\right> \nonumber \\
  &   & \ \ \ \ \ \ \ +  \int\! \frac{d\omega}{2\pi} \, \int\! dt \, e^{i\omega t}\left<Q_2(t)Q_1(0)\right>
\label{EQN:W-2Q}
\end{eqnarray}
where $W$ is the width of a single-qubit MRT peak~\cite{amin-mrt-2008,harris-mrt-2008-truncated}. If $Q_1(t)$ and $Q_2(t)$ are uncorrelated the final integrals of Eq.~(\ref{EQN:W-2Q}) will be zero, thus yielding $W_{2q}=\sqrt{2}W$.  As in the single qubit case, the fluctuation-dissipation theorem connects the peak width and displacement to the temperature of the environment $T$: $W_{2q}^2 = 2 T \epsilon_{p,2q}$~\cite{weiss-dissipative-book}. 

\begin{figure}
\includegraphics[width=3.5in]{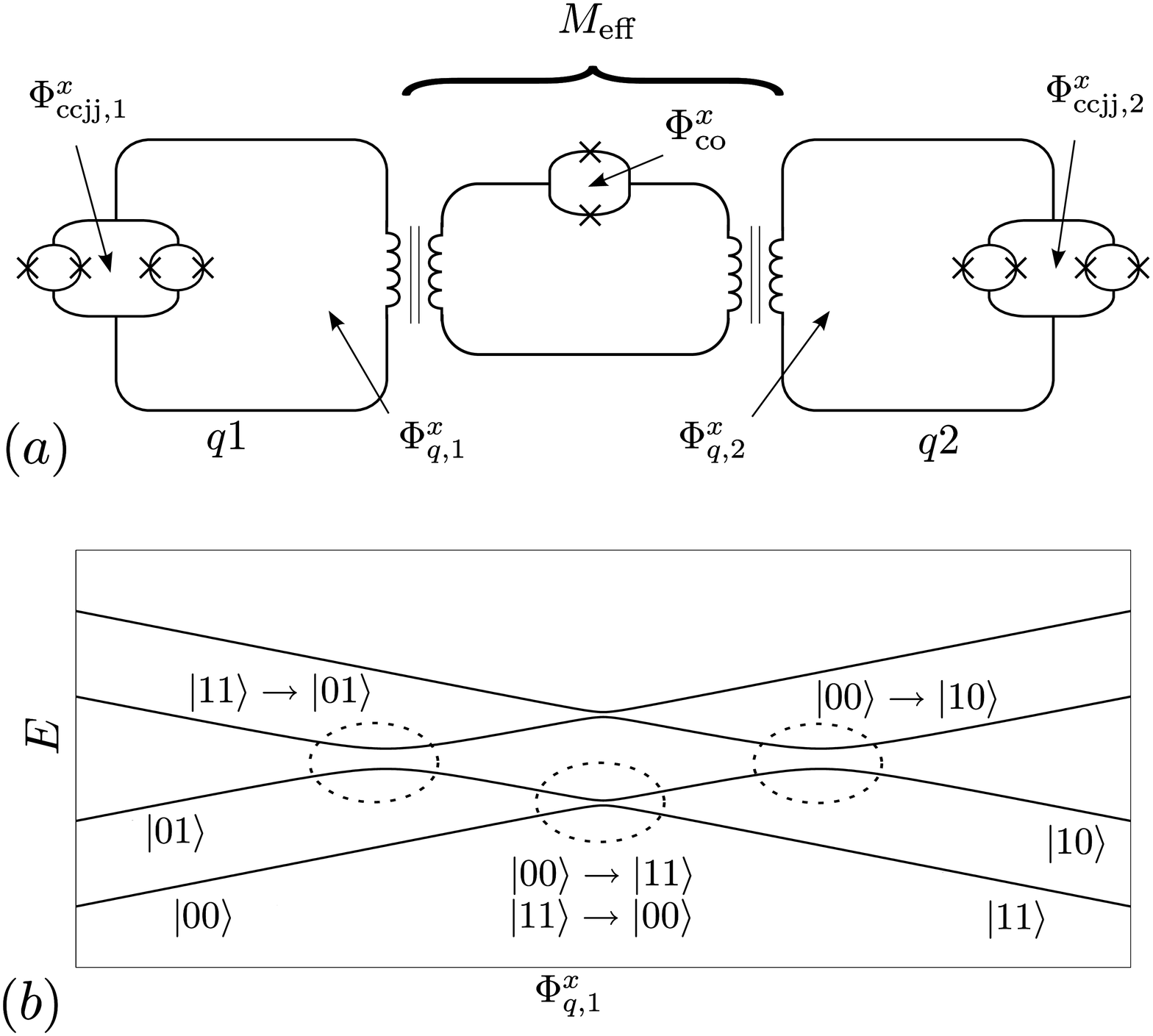}
 \caption{$(a)$ Schematic showing two compound-compound Josephson junction flux qubits and a tunable coupling element. For the experiments reported on herein, we had independent time varying control over \phicjj and $\Phi^x_q$ for each qubit. The coupling strength was tuned with a static flux bias $\Phi^x_{\rm{co}}$. $(b)$ Example eigenspectrum for a strongly FM coupled pair of flux qubits. The four lowest lying diabats are \ket{00}, \ket{11}, \ket{01}, and \ket{10}. Anticrossings that give rise to specific resonant tunneling processes are highlighted with dashed ellipses and have been denoted as $|\alpha\beta\rangle \rightarrow |\delta\gamma\rangle$.} 
  \label{FIG:schematic}
\end{figure}

We performed measurements on a chip that includes eight compound-compound Josephson junction (CCJJ) rf-SQUID flux qubits~\cite{harris-ccjj-2010-truncated} with sixteen pair wise tunable coupling elements~\cite{harris-cjc-2009-truncated}. Figure~\ref{FIG:schematic}(a) shows a simplified schematic of two qubits connected by a coupler. For further details on this circuit see Refs.~\cite{harris-ccjj-2010-truncated,harris-cjc-2009-truncated,johnson-pmm-2010-truncated,berkley-readout-2010-truncated}. The chip was manufactured on an oxidized Si wafer with Nb/Al/Al$_2$O$_3$/Nb trilayer junctions and four Nb wiring layers insulated from one another with planarized high density plasma enhanced chemical vapor deposited SiO$_2$. We mounted the chip in an Al box on the mixing chamber of a dilution refrigerator. All measurements reported herein were performed at $T = 21$ mK.

\begin{figure}[t]
  \includegraphics[width=3.25in]{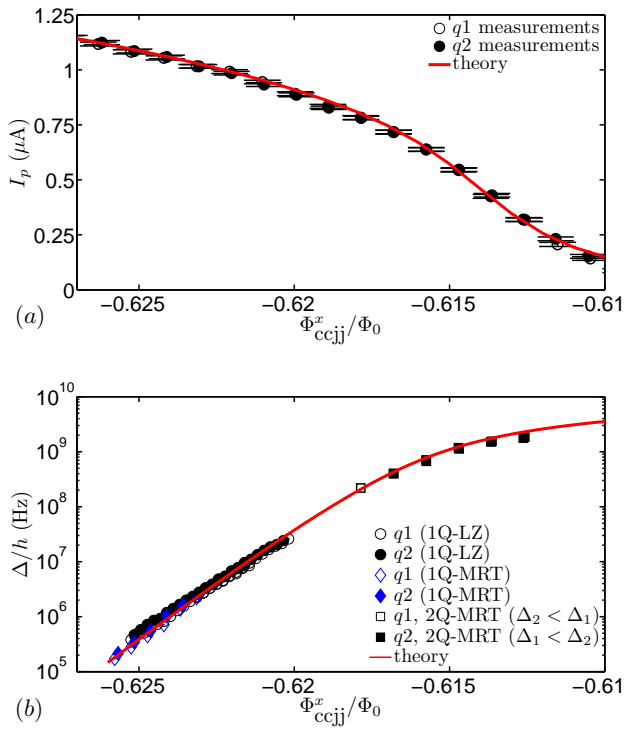}
  \caption{$(a)$ \ip versus \phicjj for $q1$ and $q2$. $(b)$ \deltaq$\ $ versus \phicjj for $q1$ and $q2$.  \deltaq$\ $ was obtained from single-qubit LZ (1Q-LZ), single-qubit MRT (1Q-MRT), and two-qubit MRT (2Q-MRT) measurements.
  \label{FIG:single-q-curves-127}}
\end{figure}

For every coupled pair of qubits in this circuit we had independent time-varying control over the CCJJ flux bias \phicjj and the qubit body flux bias $\Phi^x_q$~\cite{harris-ccjj-2010-truncated}. These signals were provided by room temperature current sources with cold filtering that limited the bandwidth to 5 MHz. The couplers provided mutual inductances $M_{\rm{eff}}$ between pairs of qubit loops and which could be tuned from $1.8$ pH to $-3.0$ pH via a static flux bias $\Phi^x_{\rm{co}}$~\cite{harris-cjc-2009-truncated}. We focus the rest of the paper on results obtained from a particular pair of qubits we call $q1$ and $q2$, which we isolated from the rest of the circuit by setting all but one of the interqubit couplers to $M_{\rm{eff}} = 0$. We have reproduced these results with the other 15 pairs of qubits on this chip.

We began our experimental investigation of this chip by calibrating all on-chip mutual inductances and qubit parameters. Reference \cite{harris-ccjj-2010-truncated} describes these calibration techniques for an identical chip. We obtained a qubit critical current $I_c = 3.38 \pm 0.01\ \mu$A, a qubit inductance $L_q = 338\pm 1$ pH, a CJJ loop inductance $L_{\rm{ccjj}} = 26\pm 1$ pH and a qubit capacitance $C = 185 \pm 5$ fF for $q1$ and $q2$. 

\begin{figure}[t]
  \includegraphics[width=3.25in]{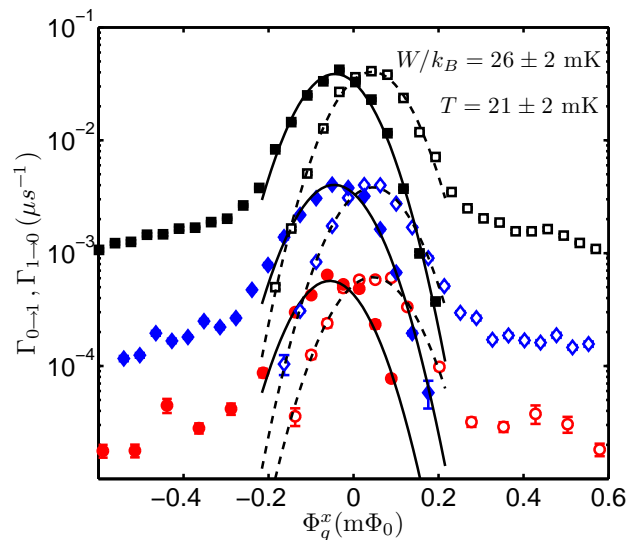}
  \caption{Example measurements of single qubit MRT rates versus $\Phi^x_q$ for $q2$. The hollow (solid) symbols are $\Gamma_{0\rightarrow1}$ ($\Gamma_{1\rightarrow0}$). Data shown are for $\Phi^x_{\rm{ccjj}}/\Phi_0 =  -0.6231, -0.6242,$ and $-0.6253$ from top to bottom, respectively. The curves are fits to Eq.~(2) of \cite{harris-mrt-2008-truncated}.
  \label{FIG:1QMRT}}
\end{figure}

After the parameter calibration noted above, we measured \ip and \deltaq$\ $ as a function of \phicjj, as summarized in Fig.~\ref{FIG:single-q-curves-127}. We measured $I_p$ by using a second qubit as a sensor of coupled flux, as described in Ref.~\cite{harris-ccjj-2010-truncated}. We measured \deltaq $\ $ via three methods. The first method used MRT between the ground and first excited states of a single qubit~\cite{harris-mrt-2008-truncated}. Figure~\ref{FIG:1QMRT} shows example single qubit MRT rate measurements for three values of \phicjj. We obtained $W/k_B = 26 \pm\ 2$ mK ($W/2I_p = 80$ $\pm\ 6\ \mu\Phi_0$) from fitting these data to Eq.~(2) of~\cite{harris-mrt-2008-truncated}. The second method involved Landau-Zener (LZ) rate measurements~\cite{johansson-lz-2009-truncated}. The bandwidth restrictions of our cold filtering limited both of the above techniques to measuring \deltaq$/h$ $\lesssim 50$ MHz. To characterize larger \deltaq, we used an alternate method that will be described below. The solid curves in Fig.~\ref{FIG:single-q-curves-127} are the theoretical predictions of the ideal CCJJ rf-SQUID Hamiltonian given in Ref.~\cite{harris-ccjj-2010-truncated} using the independently calibrated qubit parameters $I_c,\ L_q,\ L_{\rm{ccjj}}$ and $C$ cited above. The quality of the agreement between experimental data and theory justifies our identification of these devices as flux qubits.

\begin{figure}[h]
  \includegraphics[width=3.25in]{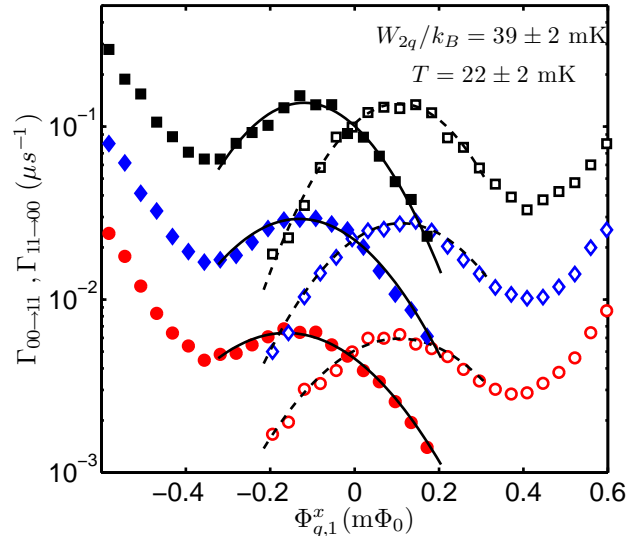}
  \caption{Example measurements of two-qubit MRT rates versus $\Phi^x_{q,1}$) for two coupled qubits ($M_{\rm{eff}} =-2.35$ pH). The hollow (solid) symbols are $\Gamma_{00\rightarrow11}$ ($\Gamma_{11\rightarrow00}$). Data are shown for both qubits biased at $\Phi^x_{\rm{ccjj}}/\Phi_0 = -0.6176, -0.6181,$ and $-0.6187$ from top to bottom, respectively. The curves are fits to Eq. (\ref{EQN:2q-mrt-eqn}).
  \label{FIG:2QMRT}}
\end{figure}



With the single qubit calibrations completed, we then turned to experiments on pairs of strongly FM coupled qubits. The two-qubit MRT experiment was performed in a manner very similar to the single qubit case~\cite{harris-mrt-2008-truncated}. We applied $\Phi^q_x = \pm$10 m$\Phi_0$ to each qubit in the presence of \phicjj = $-\Phi_0/2$, raised their tunnel barriers by ramping \phicjj from $ -\Phi_0/2$ to  $-\Phi_0$, and then waited 1 ms to ensure the two-qubit pair was in its ground state. This initialized the coupled pair in either \ket{00} or \ket{11} with certainty. Next we adjusted $\Phi^q_{x,i}$ and then simultaneously lowered the tunnel barriers of both qubits for a dwell time $\tau$ before again raising them via the individual \phicjj. We measured the loss of probability from the initial state and repeated for a range of $\tau$. The probability of the initial state as a function of $\tau$ was fit to an exponential to extract $\Gamma_{00\rightarrow11}$ or $\Gamma_{11\rightarrow00}$ depending upon the initialization.

Having individual control of \phicjj for each member of a pair allowed us to perform two-qubit MRT measurements in which we either matched $\Delta_1 = \Delta_2$ or deliberately mismatched $\Delta_1 \neq \Delta_2$. For the first set of measurements, we set $\Delta_1 = \Delta_2$ by biasing $\Phi^x_{\rm{ccjj, 1}} = \Phi^x_{\rm{ccjj, 2}}$.  Figure \ref{FIG:2QMRT} shows example measurements of $\Gamma_{00\rightarrow11}$ and $\Gamma_{11\rightarrow00}$ as a function of $\Phi^x_{q,1}$ with $\Phi^x_{q,2} = 0$ and $M_{\rm{eff}} = -2.35$ pH. The increase in rate for $|\Phi^x_{q,1}| > 0.4\ \rm{m}\Phi_0$ is due to tunneling from the initial state, either \ket{00} or \ket{11}, to \ket{01} or \ket{10}, by the processes depicted in Fig.~\ref{FIG:schematic}(b). For these experimental settings, it was predicted that these processes would peak at $\pm 0.8$ m$\Phi_0$.

\begin{figure}
  \includegraphics[width=3.5in]{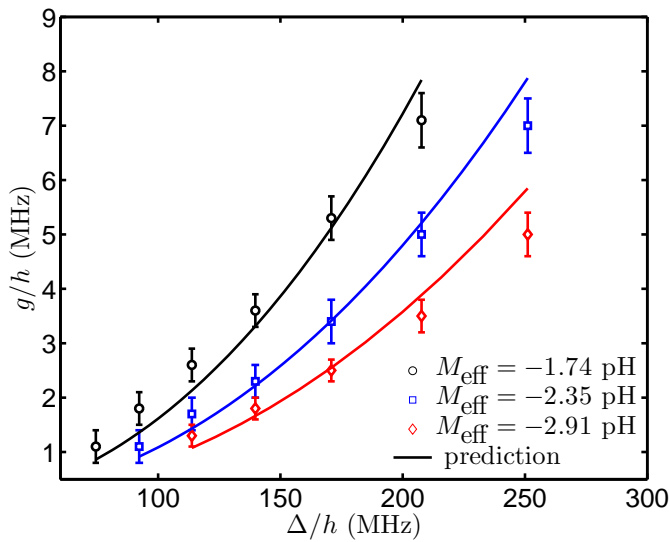}
  \caption{Two qubit gap $g$ versus $\Delta$ for a range of $M_{\rm{eff}}$. The qubits were biased at matching \phicjj. Values of $g$ were obtained from fits such as those shown in Fig.~\ref{FIG:2QMRT}. The solid curves are the theoretical predictions of from Eq.~(\ref{EQN:fullanalyticgap}). 
  \label{FIG:gap-measurements}}
\end{figure}

To extract $\epsilon_{p,2q}$ and $W_{2q}$ from data such as those in Fig.~\ref{FIG:2QMRT}, we fit the MRT rate peaks to Eq.~(\ref{EQN:2q-mrt-eqn}). For all measurements with $\Delta_1=\ \Delta_2$ we obtained $W_{2q}/k_B = 39\pm 2$~mK. The ratio $W_{2q}/W~=~1.5~\approx~\sqrt{2}$, which indicates that the environment coupled to $q1$ is uncorrelated with that coupled to $q2$. This is evidence that the source of flux noise in these qubits is local to the qubit wiring, which agrees with the conclusions of others ~\cite{yoshihara-correlated-2010}. Values of $W_{2q}$ and $\epsilon_{p,2q}$ were used to infer a temperature $T~=~W_{2q}^2/2\epsilon_{p,2q}~=~22\ \pm 2$ mK. This is consistent with $T$ as determined via single qubit MRT and with that reported by thermometry. 

Fitting MRT peaks to Eq.~\ref{EQN:2q-mrt-eqn} also allowed us to extract the two-qubit energy gap $g$. Figure \ref{FIG:gap-measurements} shows $g$ for a range of single qubit $\Delta \equiv \Delta_1 = \Delta_2$ and three different coupling strengths. The theoretical predictions were generated using Eq.~(\ref{EQN:fullanalyticgap}), where we used $\Delta$ as predicted by the theoretical curve shown in Fig. \ref{FIG:single-q-curves-127}(b) and an independent calibration of $J$. There is good agreement between the measured and predicted $g$ for different $M_{\rm{eff}}$. We conclude that $g \propto \Delta^2$ and $g \propto 1/J$, as predicted when $\Delta \ll 2|J|$.  

Besides having achieved the goal of demonstrating quantum mechanical behavior in pairs of coupled qubits, we have used two-qubit MRT measurements to complement our single qubit methods for calibrating large $\Delta$. We biased control qubit $q1$ at $\Phi^x_{\rm{ccjj}}/\Phi_0 = -0.6222$, a point at which $\Delta_1/h~=~6.0~\pm 0.2$ MHz had been independently calibrated using single qubit methods. We then targeted  \phicjj$/\Phi_0 >  -0.6222$ of the second qubit $q2$, which ensured that $\Delta_1 < \Delta_2$.  The dynamics of the coupled system were then governed by $g < \Delta_1$, which made them measureable given our experimental bandwidth. We extracted $g$ from a two-qubit MRT measurement of this mismatched pair which, along with the independently calibrated values of $\Delta_1$ and $J$, allowed us to infer $\Delta_2$ from Eq. (\ref{EQN:fullanalyticgap}). We then switched the roles of $q1$ and $q2$ to infer $\Delta_1$. The results of such experiments using $M_{\rm{eff}} = -2.91$ pH have been summarized in Fig.~\ref{FIG:single-q-curves-127}(b). Again, the experimental data agree with the predictions of the independently calibrated ideal CCJJ rf-SQUID Hamiltonian, further confirming the self-consistency of our measurements. Thus we have validated a new technique for characterizing high tunneling energies of single qubits despite the limited bandwidth of our apparatus.

{\em{Conclusions.}} Macroscopic resonant tunneling is a powerful way of characterizing single and coupled pairs of superconducting flux qubits. 
We have demonstrated that inductively coupled pairs of flux qubits behave as expected by quantum mechanics in that the two-qubit tunneling energy $g$ inferred from fitting experimental data agrees with the predictions of an effective two-level Hamiltonian for the coupled qubit system. Investigating one pair in detail for a range of $M_{\rm{eff}}$, $\Delta_1$, and $\Delta_2$ yielded $g$ that not only matched theoretical predictions, but allowed us to probe single qubit tunneling energies $0.2$~GHz~$\lesssim$~\deltaq/$h$~$\lesssim 2$ GHz without the use of microwave lines. Finally, the widths of the two-qubit MRT rate peaks were a factor of $\sqrt{2}$ larger than that of a single qubit.  It was argued that this is an indication that the environment interacting with one qubit is uncorrelated with that of the other qubit.  This latter observation implies that the source of flux noise is local to the qubit. 

We thank D. Averin, J. Hilton, P. Spear, A. Tcaciuc, F. Ciota, D. Klitz and L. Paulson for useful discussions and experimental support. 



\end{document}